\documentclass[conference]{IEEEtran}

\usepackage[noadjust]{cite}

\usepackage{amsmath,amssymb,amsfonts}
\usepackage{algorithmic}
\usepackage{graphicx}
\usepackage{textcomp}

\usepackage[dvipsnames]{xcolor}

\usepackage[hyphens,spaces,obeyspaces]{url}

\usepackage{amsmath}

\usepackage{tikz}

\usetikzlibrary{positioning}

\tikzset{basic/.style={draw,fill=blue!10,text width=1em,text badly centered}}
\tikzset{input/.style={basic,circle}}
\tikzset{weights/.style={basic,rectangle}}
\tikzset{functions/.style={basic,circle,fill=blue!10}}

\usetikzlibrary{shapes,arrows}
\tikzstyle{block} = [draw, fill=blue!20, rectangle, minimum height=3em, minimum width=6em]
\tikzstyle{sum} = [draw, fill=blue!20, circle, node distance=1cm]
\tikzstyle{input} = [coordinate]
\tikzstyle{output} = [coordinate]
\tikzstyle{pinstyle} = [pin edge={to-,thin,black}]

\newcommand\Tstrut{\rule{0pt}{2.6ex}}         
\newcommand\Bstrut{\rule[-0.9ex]{0pt}{0pt}}   

\def\BibTeX{{\rm B\kern-.05em{\sc i\kern-.025em b}\kern-.08em
    T\kern-.1667em\lower.7ex\hbox{E}\kern-.125emX}}

\makeatletter

\def\ps@IEEEtitlepagestyle{%
    \def\@oddfoot{\mycopyrightnotice}%
    \def\@evenfoot{}%
}
\def\mycopyrightnotice{%
    {\footnotesize  Sophos Technical Papers, Feb, 2019\hfill}
    \gdef\mycopyrightnotice{}
}

\title{\huge Machine Learning With Feature Selection Using Principal Component Analysis for Malware Detection: A Case Study}

\begin{document}

\author{\IEEEauthorblockN{\textbf{Jason Zhang}\textit{, Ph.D.}}
\IEEEauthorblockA{\textit{Senior Threat Researcher}}
\textit{Sophos, Abingdon OX14 3YP, U.K.}\\
jason.zhang@sophos.com
}

\thispagestyle{plain}
\pagestyle{plain}

\maketitle

\begin{abstract}
Cyber security threats have been growing significantly in both volume and sophistication over the past decade. This poses great challenges to malware detection without considerable automation. In this paper, we have proposed a novel approach by extending our recently suggested artificial neural network (ANN) based model with feature selection using the principal component analysis (PCA) technique for malware detection. The effectiveness of the approach has been successfully demonstrated with the application in PDF malware detection. A varying number of principal components is examined in the comparative study. Our evaluation shows that the model with PCA can significantly reduce feature redundancy and learning time with minimum impact on data information loss, as confirmed by both training and testing results based on around $105,000$ real-world PDF documents. Of the evaluated models using PCA, the model with $32$ principal feature components exhibits very similar training accuracy to the model using the $48$ original features, resulting in around $33\%$ dimensionality reduction and $22\%$ less learning time. The testing results further confirm the effectiveness and show that the model is able to achieve $93.17\%$ true positive rate (TPR) while maintaining the same low false positive rate (FPR) of $0.08\%$ as the case when no feature selection is applied, which significantly outperforms all evaluated seven well known commercial antivirus (AV) scanners of which the best scanner only has a TPR of $84.53\%$.
\end{abstract}


\section{Introduction}
During the past decade, machine learning (ML), deep learning (DL) or generally termed artificial intelligence (AI) have gained wide popularity with applications across industries thanks to exponential improvement in computing hardware, availability of big datasets and improved algorithms. Recent work has witnessed breakthroughs from image classification, speech recognition to robot control and autonomous driving. In the meantime, cyber security attacks have been growing significantly in both volume and sophistication over the past decade as well. Typical examples include spamming campaigns, phishing emails and application vulnerability exploits, either via targeted or non-targeted attacks. It's no surprise that AI based methods have been increasingly studied or applied in cyber security applications, varying from spam filtering \cite{Wu09,KumarGWM16C}, intrusion detection \cite{Engen10T}, malware detection \cite{Sophos17} and classification \cite{PascanuSSMT15C}. 

A crucial step in an ML workflow is feature extraction which can be hand-crafted based on human expertise, or automatically learned by training modern deep learning models such as convolutional neural networks (CNNs). It is natural to believe that more extracted features are able to provide better characterization of a learning task and more discriminating power. However, increasing the dimension of the feature vector could result in feature redundancy and noise. Redundant and irrelevant features can cause performance deterioration of an ML model with overfitting and generalization problems. Additionally, excessively increased number of features could significantly slow down a learning process. Therefore it is of fundamental importance to only keep relevant features before feeding them into an ML model, which leads to the requirement of feature selection (or feature dimensionality reduction). Feature selection can be seen as the process of identifying and removing as much of the noisy and redundant information as possible from extracted features. 

To demonstrate the effectiveness of our proposed ML approach with feature selection using PCA in malware detection, the suggested approach is evaluated with PDF malware detection as a case study.

Portable document format (PDF) is widely used for electronic documents exchange due to its flexibility and independence of platforms. PDF supports various types of data including texts, images, JavaScript, Flash, interactive forms and hyper links, etc. The popularity and flexibility of PDF also provide opportunities for hackers to carry out cyber attacks in various ways. A common PDF based attack is phish, such as PDF based order confirmation and parcel delivery notice, which typically uses social engineered texts to entice users to click phishing links embedded in PDF documents. Another typical attack using PDF is exploits which target vulnerable PDF reading applications. For example, an attack detected by Sophos targets four vulnerabilities of a popular PDF reading application \cite{Z15}: \textit{Collab.collectEmailInfo} (CVE-2007-5659), \textit{Util.printf}() (CVE-2008-2992), \textit{Collab.getIcon} (CVE-2009-0927) and \textit{Escript.api plugin media player} (CVE-2010-4091). Each of the exploits affects a particular version of the vulnerable application depending on the version installed on a victim's device.

There exist various approaches for the detection of PDF based attacks. Typical traditional methods include signature match and sandbox based analysis \cite{TzermiasSPM11C,RatanaworabhanLZ09C,WillemsHF07}. Given the popularity of ML in the past decade, it has also witnessed several ML applications in PDF malware detection, such as methods for detecting JavaScript in PDF files \cite{Wepawet, LaskovS11C}. Other learning-based approaches for general PDF malware detection include a decision tree based ensemble learning algorithm \cite{CrossM11T}, as well as methods using Naive Bayes, support vector machines (SVM) and Random Forest \cite{MaiorcaGC12,SmutzS12C,SrndicL13C,CuanDDV18T}. The pros and cons of the traditional methods and ML based approaches above are discussed in \cite{Z18C}.

In this paper, we have proposed a novel approach by extending our recently proposed MLP neural network model MLP$_{df}$ \cite{Z18C} with feature selection using PCA for malware detection. The performance is evaluated with PDF malware detection before and after applying feature selection. Feature selection with dimensionality reduction from $33\%$ up to $79\%$ is studied. Our evaluation based on real-world data shows that the model with PCA based feature selection can significantly reduce feature redundancy and learning time with minimum impact on data information loss, as confirmed by both training and testing results.

The remainder of this paper is organized as follows: In the following section, the MLP$_{df}$ model from \cite{Z18C} is introduced. Section \ref{FEng} discusses feature extraction, followed by feature selection using PCA. Then, Section \ref{results} contains evaluation results illustrating the performance of the proposed approach with comparison to other methods. Finally, Section \ref{concl} contains our conclusions and suggested future work.

\section{The MLP$_{df}$ Model}\label{MLPdf}
The MLP$_{df}$ neural network model we recently proposed in \cite{Z18C} has an input layer, an output layer, and two hidden layers between them. It is a densely connected network, which means each node in one layer fully connects to every node in the following layer. In order to carry out the supervised learning process, it is necessary to extract a digital representation $\mathbf{x}$ of a given object or event that needs to be fed into the MLP model. The learning task becomes to find a multidimensional function $\Psi(\cdot)$ which maps the input $\mathbf{x}$ to the target $\mathbf{y}$, as shown below
\begin{equation}\label{eq:ml}
\mathbf{y} \cong \Psi(\mathbf{x})
\end{equation}
where $\mathbf{x} \in \mathbb{R}^{N \times 1}$, a real-valued input feature vector $\mathbf{x} = [x_{1}, \cdots, x_{N}]^{T}$ in an $N$ dimensional feature space, with $(\cdot)^T$ denoting the transpose operation. Details on features and feature engineering will be discussed in Section \ref{FEng}. Similarly, $\mathbf{y} \in \mathbb{R}^{M \times 1}$, a real-valued target classification vector $\mathbf{y} = [y_{1}, \cdots, y_{M}]^{T}$ in an $M$ dimensional classification space. In the application of PDF based malware detection, $\mathbf{y}$ is a scalar ($M = 1$). 

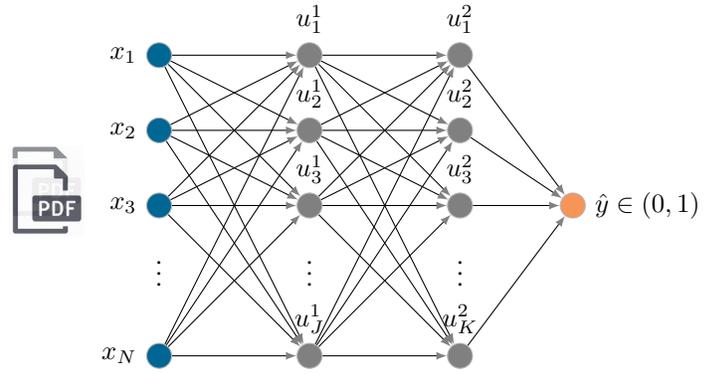
\begin{figure}[htbp]
\centering
\begin{tikzpicture}
[   cnode/.style={draw=black!30,fill=#1,minimum width=3mm,circle},
]
    \node at (0,-2.8) {\includegraphics[scale=0.15]{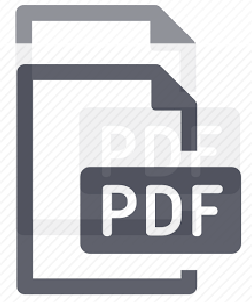}};
    \node[cnode=Peach,label=0:$\hat y \in {(0,1)}$] (s) at (7,-3) {};
    \node at (1.5,-3.8) {$\vdots$};
    \node at (3.5,-3.8) {$\vdots$};
    \node at (5.5,-3.8) {$\vdots$};
    \foreach \x in {1,...,4}
    {   \pgfmathparse{\x<4 ? \x : "N"}
        \node[cnode=MidnightBlue,label=180:$x_{\pgfmathresult}$] (x-\x) at (1.5,{-\x-div(\x,4)}) {};
        \pgfmathparse{\x<4 ? \x : "J"}
        \node[cnode=gray,label=90:$u^{1}_{\pgfmathresult}$] (p-\x) at (3.5,{-\x-div(\x,4)}) {};
        \pgfmathparse{\x<4 ? \x : "K"}
        \node[cnode=gray,label=90:$u^{2}_{\pgfmathresult}$] (z-\x) at (5.5,{-\x-div(\x,4)}) {};
        \draw [-{latex[black!50]}] (z-\x) -- node[above,sloped,pos=0.3]{} (s);
    }
    \foreach \x in {1,...,4}
    {   \foreach \y in {1,...,4}
        {   \draw [-{latex[black!50]}](x-\x) -> (p-\y);
            \draw [-{latex[black!50]}](p-\x) -> (z-\y);
        }
    }
\end{tikzpicture}
\caption{Architecture of the MLP$_{df}$ Model \cite{Z18C}}
\label{fig-mlp}
\end{figure}

Fig. \ref{fig-mlp} depicts the architecture of the MLP$_{df}$ model, where $\mathbf{x}$ and $\hat y$ denote input feature vector and trained binary output, respectively. $u^{i}_{k}$ denotes the neuron unit $k \in \mathbb{R}^{{J,K}}$ in the $i^{th}$ hidden layer, with $i = 1,2$ in this work. Each interconnection between the layers in the model is associated with a scalar weight $w_{j,k}$ which is adjusted during the training phase, where $j$ and $k$ are neuron unit indices between two consecutive layers.

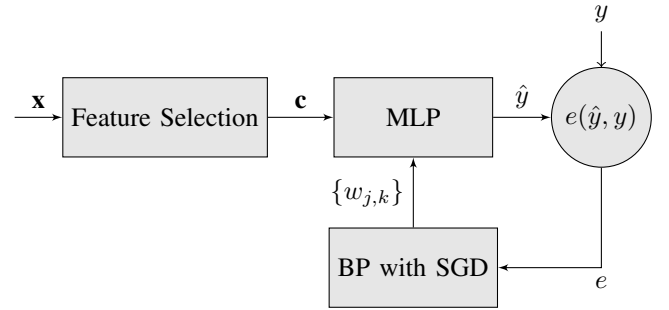
\begin{figure}[htbp]
\centering
\begin{tikzpicture}[auto, node distance=2cm,>=latex']
    \node [input, name=input] {};
    \node [block, right of=input, fill={rgb:gray,1;white,4}] (featS) {Feature Selection};
    \node [block, right of=featS, node distance=3.3cm, fill={rgb:gray,1;white,4}] (system) {MLP};
    
    \node [block, below of=system, fill={rgb:gray,1;white,4}] (sgd) {BP with SGD};

    \draw [draw,->] (input) -- node {$\textbf{x}$} (featS);
    \draw [draw,->] (featS) -- node {$\textbf{c}$} (system);

    \node [sum, right of=system, pin={[pinstyle]above:$y$},
            node distance=2.5cm, fill={rgb:gray,1;white,4}] (sigma) {$e(\hat y, y)$};
    \draw [->] (system) -- node {$\hat y$} (sigma);
    \draw [->] (sigma)  |- node {$e$}(sgd);

    \draw [->] (sgd) -- node {$\{w_{j,k}\}$} (system);
\end{tikzpicture}
\caption{MLP$_{df}$ weights update via BP algorithm with SGD search.}
\label{fig-bp}
\end{figure}

The goal of the MLP learning process is to find a set of optimal weights $\{w_{j,k}\}$ over a training dataset in order to produce the outputs as close as possible to the target values. As Fig. \ref{fig-bp} shows, the learning process with weights update is realized through the back-propagation (BP) algorithm which employs stochastic gradient decent (SGD) search through the space of possible weight values to minimize the error signal $e(\hat y, y)$ between the trained output $\hat y$ and the target value $y$. In our training datasets, malicious PDF files are labelled with $1$ and benign files are labelled with $0$. The labels are then used as our target values $\{y\}$ for the training model. More details on how the weights $\{w_{j,k}\}$ are updated during the learning process are discussed in \cite{Z18C,Mitchell97,GoodfellowBC16}. In Fig. \ref{fig-bp}, there exists a process of feature selection between the original features $\mathbf{x}$ and the MLP model. This is to remove irrelevant features and only keep the most representative feature components $\mathbf{c}$. It is an important part of the so-called feature engineering process, as discussed in the following section.

\section{Feature Engineering}\label{FEng}
Feature engineering is a crucial step in machine learning. It is to make a task easier for an ML model to learn. One should not expect an ML model to be able to learn from completely arbitrary data. In many cases, data features rather than the raw data are used as input signals for an ML model. There is no exception when dealing with the detection of PDF attacks herein. There exist hand-crafted feature engineering using specific domain knowledge and automated feature engineering such as CNN based models. In this work, the manual process of feature engineering is carried out. It includes feature extraction from raw PDF files and feature selection using PCA technique.

\begin{table}[hhtb]
\caption{Datasets and feature information}
\begin{center}
\begin{tabular}{|c|c|c|c|c|}
\hline
\multicolumn{2}{|c|}{\textbf{Training: $90000$}} & \multicolumn{2}{|c|}{\textbf{Testing: $15047$}}&\textbf{Features: $48$}\Tstrut\Bstrut\\
\hline
\textit{Benign}& \textit{Malicious}& \textit{Benign}& \textit{Malicious}& Structure, metadata\Tstrut\Bstrut \\
\cline{1-4} 
\Tstrut\Bstrut
\textit{$78684$}& \textit{$11316$}& \textit{$13101$}& $1946$ & Objects, content stats, etc. \Tstrut\Bstrut\\
\hline
\end{tabular}
\label{tbl:feat}
\end{center}
\end{table}

\subsection{Feature Extraction}\label{FE}
The information of the training and testing datasets used in this work is shown in Table \ref{tbl:feat}. An initial group of $48$ features are extracted from the datasets. In-house tools and off-the-shelf PDF parsers are used to extract the features, which comprise information from PDF structure, metadata, object characteristics as well as content statistical properties, etc. Part of the $48$ features are listed in Table \ref{tbl:flist}. As it shows, features are defined using file size, JavaScript existence, page count, object count, stream filtering, entropy value of some content. The full list of features and the procedure of extraction are not discussed in this paper due to commercial reasons. 
\begin{table}[hhtb]
\caption{Part of the extracted features}
\begin{center}
\begin{tabular}{|l|l|}
\hline
\textbf{Feature name}& \textbf{Description}\Tstrut\Bstrut \\
\hline
\Tstrut\Bstrut
\textit{F\_SIZE}& \textit{PDF file size} \Tstrut\Bstrut\\
\hline
\textit{F\_JS}& \textit{PDF with JavaScript or not} \Tstrut\Bstrut\\
\hline
\textit{F\_PGC}& \textit{Page count} \Tstrut\Bstrut\\
\hline
\textit{F\_OBJC}& \textit{Number of objects} \Tstrut\Bstrut\\
\hline
\textit{F\_FILT}& \textit{Stream filtering} \Tstrut\Bstrut\\
\hline
\textit{F\_ENTRP1}& \textit{Entropy of some content} \Tstrut\Bstrut\\
\hline
\textit{F\_ENTRP2}& \textit{Entropy of some content} \Tstrut\Bstrut\\
\hline
\textit{ $\cdots$ }& \textit{$\cdots$} \Tstrut\Bstrut\\
\hline
\end{tabular}
\label{tbl:flist}
\end{center}
\end{table}

As part of data pre-processing, all features need to be normalized before any further processes such as feature selection and training. The primary aim of normalization is to avoid large gradient updates during the SGD search and learning process. Let 
\begin{equation}\label{eq:X}
\mathbf{X} =  \big [\mathbf{x}_1, \cdots, \mathbf{x}_S \big ]^T
\end{equation}
where $\mathbf{X} \in \mathbb{R}^{S \times N}$ is a feature matrix, $\mathbf{x} \in \mathbb{R}^{N \times 1}$ a feature vector defined in (\ref{eq:ml}) and $S$ the size of a malware or benign dataset. Typically normalization is applied for each feature independently. More specifically, this is carried out along each column vector of $\mathbf{X}$. The normalized feature vector then has a standard normal (or Gaussian) distribution with $\mu=0$ and $\sigma=1$ where $\mu$ and $\sigma$ are the mean and standard deviation of the scaled feature vector, respectively.

Though we believe that the features extracted herein should possess strong discriminative power to differentiate malicious files from benign documents, it is unavoidable to have redundancy among the features. This is why a feature selection process is often applied prior to a learning process.

\subsection{Feature Selection Using PCA}\label{FSpca}
Irrelevant and redundant features can lead to an ML classifier to converge slowly and perform less well or completely fail. In this paper, the PCA technique (also known as the eigenvector regression filter or the Karhunen\texttt{-}Loeve transform \cite{GonzalezW02}) is used for dimensionality reduction, which involves zeroing out one or more of the weakest principal components, resulting in a lower-dimensional projection of the raw feature data that preserves the maximal data variance. The dimensionality reduction process is achieved through an orthogonal, linear projection operation. Without loss of generality, the PCA operation can be defined as

\begin{equation}\label{eq:pca}
	\mathbf{Y} = \mathbf{X}\mathbf{C}
\end{equation}
with $\mathbf{Y} \in \mathbb{R}^{S \times P}$ is the projected data matrix that contains $P$ principal components of $\mathbf{X}$ with $P \le N$. So the key is to find the projection matrix $\mathbf{C} \in \mathbb{R}^{N \times P}$, which is equivalent to find the eigenvectors of the covariance matrix of $\mathbf{X}$, or alternatively solve a singular value decomposition (SVD) problem for $\mathbf{X}$ \cite{GoodfellowBC16}

\begin{equation}\label{eq:pca}
	\mathbf{X} = \mathbf{U} \Sigma \mathbf{V}^T
\end{equation}
where $\mathbf{U} \in \mathbb{R}^{S \times S}$ and $\mathbf{V} \in \mathbb{R}^{N \times N}$ are the orthogonal matrices for the column and row spaces of $\textbf{X}$, and $\Sigma$ is a diagonal matrix containing the singular values, $\lambda_n$, for $n = 0, \cdots, N - 1$, non-increasingly lying along the diagonal. It can be shown \cite{GoodfellowBC16} that the projection matrix $\mathbf{C}$ can be obtained from the first $P$ columns of $\mathbf{V}$ with 

\begin{equation}\label{eq:V}
\mathbf{V} = \big [\mathbf{v}_1, \cdots, \mathbf{v}_N \big ]
\end{equation}

and

\begin{equation}\label{eq:C}
\mathbf{C} = \big [\mathbf{c}_1, \cdots, \mathbf{c}_P \big ]
\end{equation}
where $\mathbf{v}_n \in \mathbb{R}^{N \times 1}$ is the $n^{th}$ right singular vector of $\mathbf{X}$, and $\mathbf{c}_n = \mathbf{v}_n$. 

\begin{figure}[htbp]
\centerline{\includegraphics[scale=0.6]{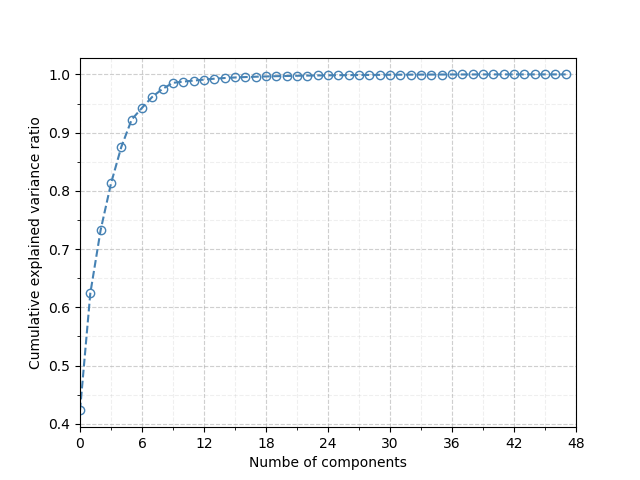}}
\caption{Cumulative explained variance ratio over components.}
\label{fig:Rcev}
\end{figure}

In fact, the singular values contained in $\Sigma$ in (\ref{eq:pca}) are the standard deviations of $\mathbf{X}$ along the principal directions in the space spanned by the columns of $\mathbf{C}$ \cite{GoodfellowBC16}. Therefore, $\lambda^2_n$ becomes the variance of $\mathbf{X}$ projection along the $n^{th}$ principal component direction. It is believed that variance can be explained as a measurement of how much information a component contributes to the data representation. One way to examine this is to look at the cumulative explained variance ratio of the principal components, given as
\begin{equation}\label{eq:Rcev}
R_{cev} = \frac{\sum_{n=1}^P\lambda^2_n}{\sum_{n=1}^N\lambda^2_n}
\end{equation}
and illustrated in Fig. \ref{fig:Rcev}. It indicates that keeping only a few principal components could retain over $90\%$ of the full variance or information of $\mathbf{X}$. As a comparative study, a varying number of principal components has been used and examined in the following evaluation section.

\section{Evaluation Results}\label{results}
The evaluation is based on the datasets described in Table \ref{tbl:feat}, which comprises $105,000$ real-world benign and malicious PDF documents collected from Sophos database. The malicious PDF documents were mostly collected over the few months up to March 2018 while the benign dataset has a longer timespan. 

It is of fundamental importance to tackle the overfitting problem in any ML process, there is no exception in our application. A trained ML algorithm must perform well on new data which was never seen during training process. Typical techniques to mitigate overfitting include \emph{Batch normalization} and \emph{Dropout} \cite{Mitchell97,GoodfellowBC16}. 
To make the evaluated models learn better during training and generalize well on new data, batch normalization with a batch size of $64$ data points is applied to the input of each layer after the input layer. This is to re-scale the input batch to have zero mean and unit variance, similar to the feature normalization discussed in Section \ref{FE}. Similarly, a dropout rate of $0.15$ is used during training process, which leads $15\%$ of each hidden layer outputs to be zeroed out before feeding into next layer. In addition, around $20\%$ of the training dataset is used as the validation dataset to help detect overfitting and perform model selection during the learning process. 

\begin{figure}[htbp]
\centerline{\includegraphics[scale=0.6]{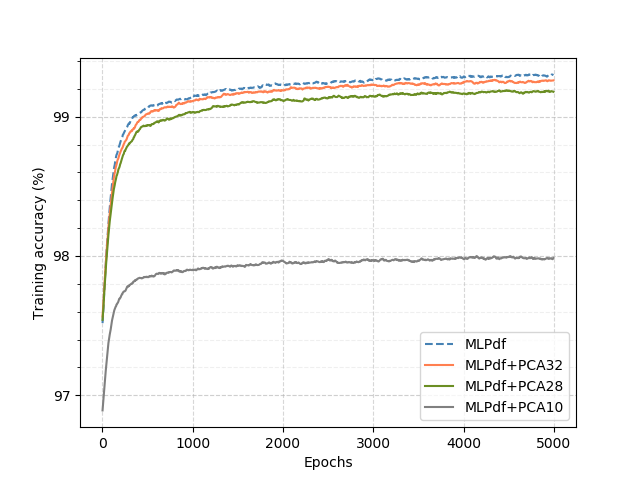}}
\caption{Training results comparison with/out feature selection.}
\label{fig:MLPdfPCA}
\end{figure}

In our previous work \cite{Z18C}, the input layer of the MLP$_{df}$ model has $48$ nodes corresponding to the number of original features used, and the output layer has a single Sigmoid (binary) probability output with values in the range of $(0, 1)$. There are two hidden layers with $72$ neurons each. In this evaluation, the number of input nodes for the models with PCA based feature selection will be dependent on the selected number of principal feature components, the rest of the model settings remains the same as the MLP$_{df}$ model. A varying number of principal components is used in the comparative study (resulting in dimensionality reduction of $79\%$, $41\%$ and $33\%$), and we term the corresponding models as MLP$_{df}+$PCA$10$, MLP$_{df}+$PCA$28$ and MLP$_{df}+$PCA$32$, respectively. These models are trained with $5000$ \emph{epoch}s each. An \emph{epoch} refers to a complete training cycle and many \emph{epoch}s are needed in order to accomplish an ANN training task. The training results are compared in Fig. \ref{fig:MLPdfPCA}. As the figure shows, all models quickly reach over $97\%$ accuracy after a small amount of training epochs, then the accuracies continue to improve with relatively slow convergence rates. It indicates that using the first $10$ principal feature components could achieve around $98\%$ training accuracy after $3000$ Epochs, this is consistent with the observation of the cumulative explained variance ratio shown in Fig. \ref{fig:Rcev}. It is no surprise that the model with $28$ principal feature components performs even better. As the figure shows, the best model using PCA is MLP$_{df}+$PCA$32$ which exhibits excellent accuracy similar to the original MLP$_{df}$ model while with around $33\%$ dimensionality reduction and $22\%$ less learning time.

\begin{figure}[htbp]
\centerline{\includegraphics[scale=0.6]{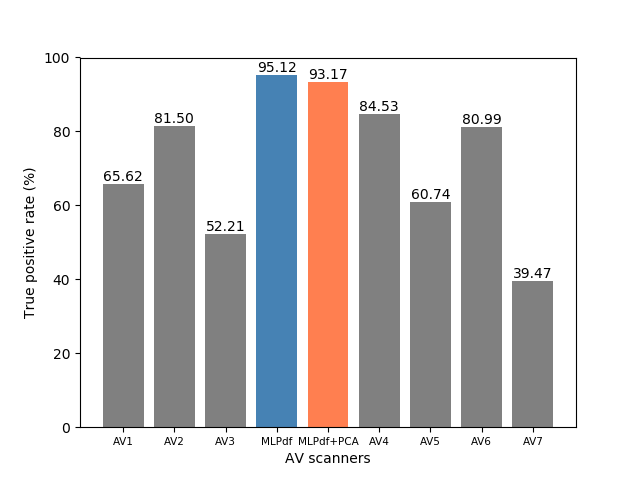}}
\caption{Testing results between MLP$_{df}$ and major AV scanners.}
\label{fig:av}
\end{figure}
To further examine the performance consistence and generalization of the model using PCA based feature selection, the model MLP$_{df}+$PCA$32$ is tested with the testing dataset shown in Table \ref{tbl:feat}, and compared with the original MLP$_{df}$ model as well as other seven major commercial AV scanners. The comparison results are shown in Fig. \ref{fig:av} where AV$1$ to AV$7$ denote the corresponding commercial AV scanners. The best result is from MLP$_{df}$ which achieves an excellent $95.12\%$ TPR while maintaining a very low FPR of $0.08\%$, as discussed in \cite{Z18C}. This is closely followed by the proposed MLP$_{df}+$PCA$32$ which manages to reach $93.17\%$ TPR with the same low FPR of $0.08\%$ maintained. Both models significantly outperform all commercial scanners. The best commercial scanner (denoted as AV4) only has a TPR of $84.53\%$.

It is worth pointing out that the evaluated seven commercial scanners perform well with zero or very low FPR with the benign testing dataset as well. One possible reason could be that our benign files are mostly collected from Sophos clean PDF documents database which has a relatively longer timespan, and the majority of them might have been already known to the commercial scanners as well. 

\section{Conclusion}\label{concl}
In this paper, we have proposed a novel approach by extending our recently proposed ANN model MLP$_{df}$ in \cite{Z18C} with feature selection using PCA technique for malware detection. As a case study, the effectiveness of the approach has been successfully demonstrated with the application in PDF malware detection. Our evaluation shows that the model with PCA can significantly reduce feature redundancy with minimum impact on data information loss, as confirmed by both training and testing results based on around $105,000$ real-world PDF documents. More specifically, a comparative study has been carried out for the model with a varying number of selected principal feature components. Of the evaluated models using PCA, the model with $32$ principal feature components, termed MLP$_{df}+$PCA$32$, exhibits very similar training accuracy to the MLP$_{df}$ model while with around $33\%$ dimensionality reduction and $22\%$ less learning time. The testing results further confirm the effectiveness and show that the proposed MLP$_{df}+$PCA$32$ model is able to achieve $93.17\%$ TPR while maintaining the same low FPR of $0.08\%$ as the case with MLP$_{df}$. On the other hand, the best commercial scanner (denoted as AV4) only manages to have a TPR of $84.53\%$. As previously pointed out, the commercial scanners perform well on the benign testing files as well with zero or very low FPR. It will be interesting to compare how the ANN based models and other commercial scanners perform with larger datasets in our future work, particularly adding more recent PDF documents to the benign corpus. Given the fact that PCA is restricted to a linear transformation, it is worth exploring non-linear feature selection techniques such as autoencoders.


\bibliographystyle{IEEEtran}

\begin{thebibliography}{10}
\providecommand{\url}[1]{#1}
\csname url@samestyle\endcsname
\providecommand{\newblock}{\relax}
\providecommand{\bibinfo}[2]{#2}
\providecommand{\BIBentrySTDinterwordspacing}{\spaceskip=0pt\relax}
\providecommand{\BIBentryALTinterwordstretchfactor}{4}
\providecommand{\BIBentryALTinterwordspacing}{\spaceskip=\fontdimen2\font plus
\BIBentryALTinterwordstretchfactor\fontdimen3\font minus
  \fontdimen4\font\relax}
\providecommand{\BIBforeignlanguage}[2]{{%
\expandafter\ifx\csname l@#1\endcsname\relax
\typeout{** WARNING: IEEEtran.bst: No hyphenation pattern has been}%
\typeout{** loaded for the language `#1'. Using the pattern for}%
\typeout{** the default language instead.}%
\else
\language=\csname l@#1\endcsname
\fi
#2}}
\providecommand{\BIBdecl}{\relax}
\BIBdecl

\bibitem{Wu09}
C.-H. Wu, ``{Behavior-based spam detection using a hybrid method of rule-based
  techniques and neural networks},'' \emph{Expert Systems with Applications},
  vol. 36(3), pp. 4321--4330, 2009.

\bibitem{KumarGWM16C}
S.~Kumar, X.~Gao, I.~Welch, and M.~Mansoori, ``{A Machine Learning Based Web
  Spam Filtering Approach},'' in \emph{Proceedings of IEEE 30th International
  Conference on Advanced Information Networking and Applications},
  Crans-Montana, Switzerland, 2016.

\bibitem{Engen10T}
V.~Engen, ``Machine learning for network based intrusion detection,'' Ph.D.
  dissertation, Bournemouth University, 2010.

\bibitem{Sophos17}
Sophos, ``{Sophos Unmatched next-gen endpoint protection: Intercept-X},''
  \url{https://www.sophos.com/en-us/products/intercept-x.aspx}, accessed:
  2018-03.

\bibitem{PascanuSSMT15C}
R.~Pascanu, J.~W. Stokes, H.~Sanossian, M.~Marinescu, and A.~Thomas, ``{Malware
  classification with recurrent networks},'' in \emph{Proceedings of 2015 IEEE
  International Conference on Acoustics, Speech and Signal Processing (ICASSP},
  Brisbane, Queensland, Australia, 2015.

\bibitem{Z15}
J.~Zhang, ``{Make ``Invisible" Visible - Case Studies in PDF Malware},'' in
  \emph{Proceedings of Hacktivity 2015}, Budapest, Hungary, 2015.

\bibitem{TzermiasSPM11C}
Z.~Tzermias, G.~Sykiotakis, M.~Polychronakis, and E.~P. Markatos, ``{Combining
  Static and Dynamic Analysis for the Detection of Malicious Documents},'' in
  \emph{Proceedings of the fourth Workshop on European Workshop on System
  Security}, Salzburg, Austria, 2011.

\bibitem{RatanaworabhanLZ09C}
P.~Ratanaworabhan, B.~Livshits, and B.~Zorn, ``{NOZZLE: A Defense Against
  Heapspraying Code Injection Attacks},'' in \emph{Proceedings of the 18th
  conference on USENIX security symposium}, Berkeley, CA USA, 2009.

\bibitem{WillemsHF07}
C.~Willems, T.~Holz, and F.~Freiling, ``{Toward Automated Dynamic Malware
  Analysis Using CWSandbox},'' \emph{IEEE Security \& Privacy}, vol. 5(2),
  2007.

\bibitem{Wepawet}
Wepawet, \url{http://wepawet.iseclab.org/}, accessed: 2018-03.

\bibitem{LaskovS11C}
P.~Laskov and N.~Srndic, ``{Static Detection of Malicious JavaScript-Bearing
  PDF Documents},'' in \emph{Proceedings of the 27th Annual Computer Security
  Applications Conference}, Orlando, Florida USA, 2011.

\bibitem{CrossM11T}
J.~S. Cross and M.~A. Munson, ``Deep pdf parsing to extract features for
  detecting embedded malware,'' Sandia National Laboratories, Albuquerque, New
  Mexico 87185 and Livermore, CA 94550, Tech. Rep. SAND2011-7982, Sep. 2011.

\bibitem{MaiorcaGC12}
D.~Maiorca, G.~Giacinto, and I.~Corona, ``{Machine Learning and Data Mining in
  Pattern Recognition},'' in \emph{volume 7376 of Lecture Notes in Computer
  Science}, Springer Berlin / Heidelberg, 2012.

\bibitem{SmutzS12C}
C.~Smutz and A.~{Stavrou}, ``{Malicious PDF Detection using Metadata and
  StructuralFeatures},'' in \emph{Proceedings of the 28th Annual Computer
  Security Applications Conference}, Orlando, Florida USA, 2012.

\bibitem{SrndicL13C}
N.~Srndic and P.~Laskov, ``{Detection of Malicious PDF Files Based on
  Hierarchical Document Structure},'' in \emph{Proceedings of the 20th Annual
  Network \& Distributed System Security Symposium}, San Diego, CA USA, 2013.

\bibitem{CuanDDV18T}
B.~Cuan, A.~Damien, C.~Delaplace, and M.~Valois, ``{Malware Detection in PDF
  Files Using Machine Learning},'' REDOCS, Tech. Rep. Rapport LAAS No. 18030,
  Feb. 2018.

\bibitem{Z18C}
J.~Zhang, ``{MLP$_{df}$: An Effective Machine Learning Based Approach for PDF
  Malware Detection},'' in \emph{Black Hat USA 2018}, Las Vegas, NV, USA, 2018.

\bibitem{Mitchell97}
T.~Mitchell, \emph{{Machine Learning}}.\hskip 1em plus 0.5em minus 0.4em\relax
  McGraw Hill, 1997.

\bibitem{GoodfellowBC16}
I.~Goodfellow, Y.~Bengio, and A.~Courville, \emph{{Deep Learning}}.\hskip 1em
  plus 0.5em minus 0.4em\relax The MIT Press, 2016.

\bibitem{GonzalezW02}
R.~C. Gonzalez and R.~E. Woods, \emph{{Digital Image Processing, 2nd
  Edition}}.\hskip 1em plus 0.5em minus 0.4em\relax AddisonWesley, 2002.

\end{thebibliography}
\end{document}